# Preserving unitarity for overlapping multichannel states: Breit-Wigner resonances vs. K matrix – comparison, advantages and disadvantages

**V. Henner**[1,2], **T. Belozerova**[1]

[1]Department of Theoretical Physics, Perm State University, 614990 Perm, Russia

[2]Department of Physics and Astronomy, University of Louisville, KY, 40292, US

## Abstract

The K-matrix method is often used to describe overlapping resonances. It guarantees the unitarity of the scattering matrix but its parameters are not resonances masses and widths. It is also unclear how to separate resonant and background contributions and to describe background in terms of phase shifts.

The Breit-Wigner (BW) approach operates with parameters having direct experimental meaning but a simple sum of the BW functions is not unitary. We show how to construct the unitary *S*-matrix by taking into account the interference of the BW functions. The method is simple and straightforward, background can be added to resonance amplitudes in the standard quantum mechanics form. In examples we give a comparison between the K-matrix and unitary BW-approaches.

## 1. Introduction

The Breit-Wigner expression [1] presents partial amplitudes in the form which directly contains masses and widths of resonances. BW function for one resonance satisfies the unitary condition, the problem arises when one needs to construct the unitary *S*-matrix for several resonances with the same quantum numbers.

A scattering operator connecting an initial and a final states, $S_{if} = \langle f|S|i\rangle$, is unitary

$$SS^{\dagger} = I \tag{1}$$

and from time-reversal invariance follows $S_{if} = S_{fi}$.

Results of any *single* resonances analysis ultimately appeal to the BW function explicitly containing the resonance parameters (we start with no background situation):

$$S_{ij} = \delta_{ij} + 2i\frac{\sqrt{\Gamma_{ri}\Gamma_{rj}}/2}{E_r - E - i\Gamma_r/2}, \qquad \Gamma_r = \sum_i^M \Gamma_{ri} \tag{2}$$

($M$ is a number of channels), or in variable $s = E^2$,

$$S_{ij} = \delta_{ij} + 2i \frac{m_r \sqrt{\Gamma_{ri}\Gamma_{rj}}}{m_r^2 - s - i m_r \Gamma_r}. \quad (3)$$

The unitarity of these expressions along with a form of the wave function of an unstable state, $\psi(t) = \psi(0) e^{-i(E_r - i\Gamma_r/2)t/\hbar}$, for which $|\psi(t)|^2 = |\psi(0)|^2 e^{-t/\tau}$ ($\tau = \hbar / \Gamma_r$ is time life of a resonance), are the base for partial amplitudes presentation in the BW form.

A simple sum of BW functions violates unitarity. An idea of writing in a general way the *S*-matrix as a sum of resonant terms which has to satisfy the unitarity constraints is due to work [2]. We demonstrate that this can be achieved in the form:

$$S_{ij} = \delta_{ij} + 2i \sum_{r=1}^{N} e^{i\varphi_{ij}^r} \frac{m_r \sqrt{\Gamma_{ri}\Gamma_{rj}}}{m_r^2 - s - i m_r \Gamma_r}, \quad (4)$$

where the interference between the resonant states is taken into account by phases $\varphi_{ij}^r$. In the original work [2] this expression was written with complex numerators (residues) instead of such phases – both forms are equivalent. This scheme was realized in work [3] for two resonances with constant widths.

To consider *energy-depending widths* and *threshold effects*, *S*-matrix should be taken in the form

$$S_{ij} = \delta_{ij} + 2iF_{ij} = \delta_{ij} + 2i\sqrt{\rho_i} T_{ij} \sqrt{\rho_j}, \quad (5)$$

where $T_{ij}$ are transition amplitudes, $\rho_i$ - phase-space factors. The conditions which unitarity imposes on widths $\Gamma_{ri}(s)$ and (constant) phases $\varphi_{ij}^r$ are formulated in section 2.

In the *K*-matrix approach [4]

$$S = \frac{I + i\rho K}{I - i\rho K}. \quad (6)$$

From $SS^\dagger = I$ and time-reversal invariance it follows that $K$ is real and symmetric operator. For the transition operator, from (6) it follows

$$T = (I - i\rho K)^{-1} K. \quad (7)$$

For elastic scattering (one channel), $S = e^{2i\delta}$ gives $K = \tan\delta$, thus a resonance in a partial amplitude at $\delta = \pi/2$ can be associated with $K$ taken in a pole form. A single pole parametrization

$$K = \frac{m_1 \Gamma_1}{m_1^2 - s} \quad (8)$$

leads to the standard BW function for the partial amplitude:

$$F = \frac{\rho m_1 \Gamma_1}{m_1^2 - s - i\rho m_1 \Gamma_1}, \quad (9)$$

thus $m_1$ and $\Gamma_1$ in (8) are mass and width of a resonance in this simple situation.

For several poles and channels, a common parametrization is [5]

$$K_{ij} = \sum_{\alpha} \frac{m_\alpha \Gamma^0_\alpha \gamma_{\alpha i} \gamma_{\alpha j}}{m_\alpha^2 - s} . \tag{10}$$

For $N$ poles and $M$ open channels, the number of free parameters in the $K$-matrix method is $N(M+1)$ [6]. Parameter $m_\alpha$ is said to be the "nominal" mass, $\gamma_{\alpha i}$ are called the coupling constants of the state $\alpha$ to the decay channel $i$. To enhance the similarity with the BW formula, they are usually normalized, $\sum_i \gamma_{\alpha i}^2 = 1$, and $\gamma_{\alpha i}$ are considered as the branching ratios. All these statements are based on comparison with the BW expression for an isolated resonance.

The example of two states and one channel allows to see the relationship between the K- and the BW -methods and the appearance of relative phases in (4). With

$$K = \frac{\gamma_1^2}{m_1 - E} + \frac{\gamma_2^2}{m_2 - E} \tag{11}$$

(here and below we use variable $E$ and omit $\rho$ only to simplify formulas illustrating a comparison between the two methods) we obtain the scattering amplitude

$$F = \frac{\gamma_1^2 m_2 + \gamma_2^2 m_1 - E\left(\gamma_1^2 + \gamma_2^2\right)}{E^2 - E\left[m_1 + m_2 - i\left(\gamma_1^2 + \gamma_2^2\right)\right] + m_1 m_2 - i\left(\gamma_1^2 m_2 + \gamma_2^2 m_1\right)} . \tag{12}$$

Denoting complex roots of the denominator as $\mu_1$ and $\mu_2$, $F$ can be written as a sum of two BW functions:

$$F = \frac{A_1}{\mu_1 - E} + \frac{A_2}{\mu_2 - E} , \tag{13}$$

where $A_{1,2}$ are complex values:

$$A_1 = \frac{\mu_1\left(\gamma_1^2 + \gamma_2^2\right) - \gamma_1^2 m_2 - \gamma_2^2 m_1}{\mu_2 - \mu_1}, \quad A_2 = \frac{\mu_2\left(\gamma_1^2 + \gamma_2^2\right) - \gamma_1^2 m_2 - \gamma_2^2 m_1}{\mu_1 - \mu_2}. \tag{14}$$

Real parts of $\mu_{1,2}$ give energies (masses) of resonances, imaginary parts – their widths:

$$E_{1,2} = \frac{m_1 + m_2}{2} \pm \frac{1}{2\sqrt{2}} \sqrt{\left(m_1 - m_2\right)^2 - \left(\gamma_1^2 + \gamma_2^2\right)^2 + W} , \tag{15a}$$

$$2\Gamma_{1,2} = \left(\gamma_1^2 + \gamma_2^2\right) \pm \frac{\sqrt{2}\left(m_1 - m_2\right)\left(\gamma_1^2 - \gamma_2^2\right)}{\sqrt{\left(m_1 - m_2\right)^2 - \left(\gamma_1^2 + \gamma_2^2\right)^2 + W}} , \tag{15b}$$

where

$$W = \sqrt{\left[\left(m_1 - m_2\right)^2 + \left(\gamma_1^2 + \gamma_2^2\right)^2\right]^2 + 16\left(m_1 - m_2\right)^2 \gamma_1^2 \gamma_2^2} . \tag{15c}$$

Thus, the K-matrix method gives for the amplitude $F$ a sum of BW functions with *complex residues*, in other words with *relative phase* – same as in (4).

A simple formula for the phase can be found. For

$$F=\sum_{n=1}^{2}\frac{\alpha_n/2}{E_n-E-i\Gamma_n/2} \tag{16}$$

$S$ is unitary if

$$\frac{\alpha_1}{\alpha_2}=\frac{\Gamma_1}{\Gamma_2}e^{2i\varphi},\quad \text{where}\quad \varphi=-\arctan\frac{\Gamma_1+\Gamma_2}{2(E_1-E_2)}. \tag{17}$$

The fact that amplitude $F$ in the $K$-method can be presented as a sum of two standard BW functions (without relative phases) if $m_1$ and $m_2$ are *very separated* is commonly considering as the justification that the $K$-matrix pole parameters are close to the physical resonance ones (when $|m_1-m_2|\gg\gamma_1^2+\gamma_2^2$, then $E_i=>m_i$, $\Gamma_i=>2\gamma_i^2$). But good separation is rather subjective argument, the unitarity is substantially violated even when resonances are rather far apart, like in Fig.1.

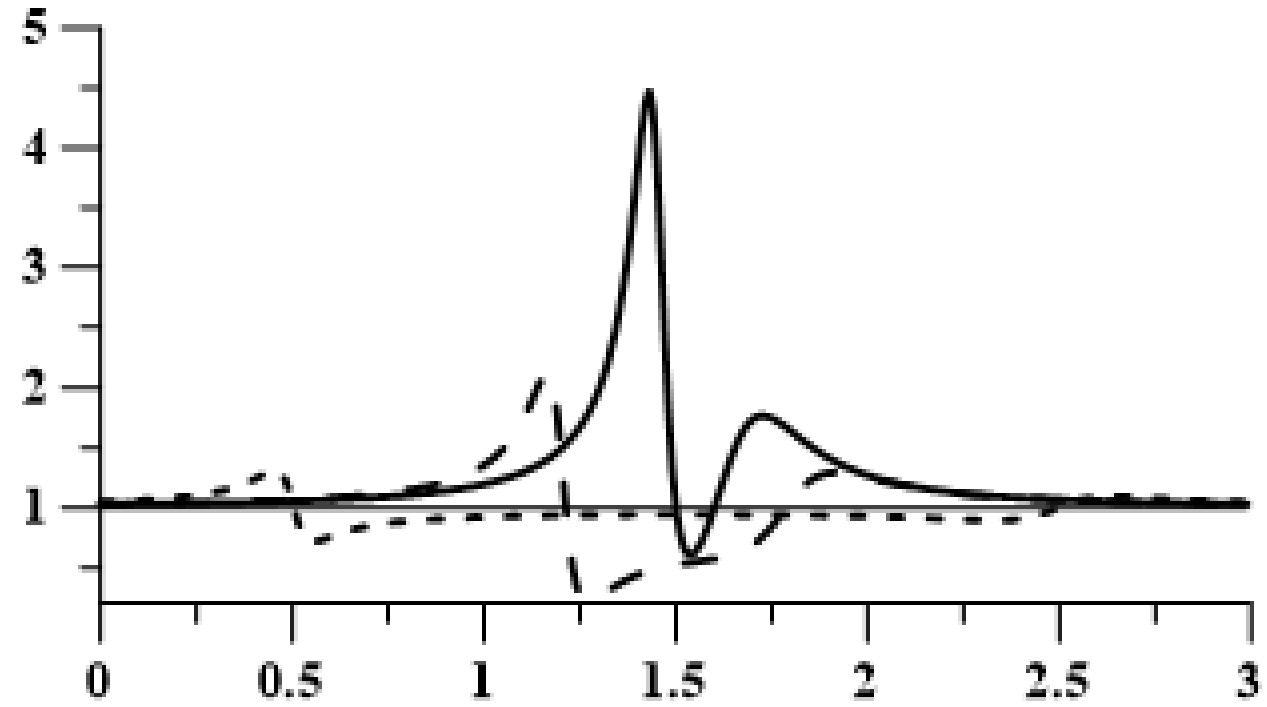

Fig. 1. Plot of $|S(E)|^2$ for a sum of two standard BW functions. Dotted line – $E_1=0.5$, $\Gamma_1=0.1$, $E_2=2.5$, $\Gamma_2=0.3$; dashed line – $E_1=1.2$, $\Gamma_1=0.1$, $E_2=1.8$, $\Gamma_2=0.3$; solid line – $E_1=1.45$, $\Gamma_1=0.1$, $E_2=1.65$, $\Gamma_2=0.3$ (all in GeV).

## 2. Brief description of the unitary BW method

Here we briefly describe the method to construct preserving unitarity $S$-matrix. The regular procedure is presented in Appendix.

The unitarity $S^\dagger S=I$ for $S=I+2iF=I+2i\sqrt{\rho}T\sqrt{\rho}$ gives

$$T^\dagger-T=iT^\dagger\rho T. \tag{18}$$

The unitarization procedure becomes technically simpler if, following the idea of works [2,3], to consider *vectors of partial widths*, $\vec{g}_r$, related with partial widths, with the components $g_{ri}=e^{i\varphi_{ri}}|g_{ri}|$, $i=1,...,M$, $r=1,...,N$, thus

$$S_{ij} = \delta_{ij} + 2i\sqrt{\rho_i(s)}\left(\sum_{r=1}^{N} e^{i\varphi_{ij}^{(r)}} \frac{m_r\Gamma_r^0 |g_{ri}|\cdot|g_{rj}|}{m_r^2 - s - im_r\Gamma_r(s)}\right)\sqrt{\rho_j(s)}, \quad (19)$$

or

$$T_{ij}(s) = \sum_{r=1}^{N} \frac{m_r\Gamma_r^0 g_{ri} g_{rj}}{m_r^2 - s - im_r\Gamma_r(s)} \quad (20)$$

(here $\varphi_{ij}^r = \varphi_{ri} + \varphi_{rj}$; factors $\Gamma_r^0$ are introduced to deal with dimensionless vectors $g_{ri}$, their values can be taken as $\Gamma_r^0 = \Gamma_r(m_r^2)$). Expression (19) is directly *T*-invariant.

The poles of the *T*-matrix correspond to the eigenvalues of the operator $M - i\Gamma/2$ and can be identified with the physical particles [6].

The branching ratio of decay of resonance *r* in channel *i* is

$$B_{ri} = \frac{|g_{ri}|^2}{|\vec{g}_r|^2} = \frac{|g_{ri}|^2}{\sum_{k=1}^{M} |g_{rk}|^2}. \quad (21)$$

The partial widths are $\Gamma_{ri} = \Gamma_r B_{ri}$. Total and partial widths can depend on energy, $\Gamma_r(s) = \sum_i^M \Gamma_{ri}(s)$.

Formulas in the rest of this section become technically simpler if to use variable *E* rather than $s = E^2$, all the expressions and algorithms can be rewritten with $s$. For

$$T_{ij} = \sum_{r=1}^{N} \frac{\Gamma_r^0 g_{ri} g_{rj}/2}{m_r - E - i\Gamma_r(E)/2} \quad (22)$$

let us formulate the conditions that should be imposed on vectors $\vec{g}_r$ to keep $S_{ij}(E)$ unitary. From (18) we have

$$i\sum_{r=1}^{N}\left[\frac{g_{ri}^* g_{rj}^*}{m_r - E + i\Gamma_r(E)/2} - \frac{g_{ri} g_{rj}}{m_r - E + i\Gamma_r(E)/2}\right] + \\ + \sum_{l=1}^{N}\sum_{r=1}^{N} V_{lr} \frac{g_{ri} g_{lj}^*}{\left(m_r - E + i\Gamma_r(E)/2\right)\left(m_r - E + i\Gamma_r(E)/2\right)} \equiv 0 \quad (23)$$

with the notation

$$V_{lr} = \sqrt{\Gamma_l^0 \Gamma_r^0} \sum_{k=1}^{M} \rho_k(E) g_{lk}^* g_{rk}. \quad (24)$$

The constraints (23) are complicated non-linear conditions. We suggest the way which allows to overcome substantial technical difficulties which previously restricted the approach to maximum two [3] or three [7] resonances even for $\rho_k \equiv 1$. This is a construction of vectors $\vec{g}_r = \vec{g}_r^x + i\vec{g}_r^y$ in the way that their imaginary parts $\vec{g}_r^y$ are combinations of their real parts $\vec{g}_r^x$: $\vec{g}_r^y = U\vec{g}_r^x$, or

$$\vec{g}_r^y = u_{r1}\vec{g}_1^x + u_{r2}\vec{g}_2^x + \ldots + u_{rN}\vec{g}_N^x. \tag{25}$$

$U$ is real anti-symmetric matrix which has a very simple form for any particular $N$ and $M$ (see Appendix). Then, instead of trying to find all the components of vectors $\vec{g}_r$, we find only their real parts and obtain their imaginary parts using matrix $U$. Immediately highlight that the number of free parameters is the *same as in the K-matrix method*, $N(M+1)$.

When $\left|m_r - m_{r'}\right| \gg \Gamma_r + \Gamma_{r'}$, matrix elements $u_{rk} \to 0$ and vectors $\vec{g}_r$ are getting real and orthogonal, $\left(\vec{g}_r, \vec{g}_q\right) = 0$.

The constraints on vectors $\vec{g}_r$ are the following:

$$\sum_{k=1}^{M} \Gamma_r^0 \rho_k(E) \,|\, g_{rk} \,|^2 = -\frac{1}{S}\left[S + 2Q_r\right]\Gamma_r, \tag{26}$$

$$\sum_{k=1}^{M} \Gamma_r^0 \rho_k(E) \mathrm{Re}\left(g_{qk}^* g_{rk}\right) = -\frac{1}{S}\left[2F_{qr}\left(m_q - m_r\right) - iG_{qr}\left(\Gamma_q + \Gamma_r\right)\right], \tag{27}$$

$$\sum_{k=1}^{M} \Gamma_r^0 \rho_k(E) \mathrm{Im}\left(g_{qk}^* g_{rk}\right) = -\frac{1}{S}\left[2G_{qr}\left(m_q - m_r\right) + iF_{qr}\left(\Gamma_q + \Gamma_r\right)\right], \tag{28}$$

$$r = 1, \ldots, N; \quad q = r+1, \ldots, N.$$

Constant coefficients $S$, $Q_r$, $F_{qr}$, $G_{qr}$ are determined via the elements of matrix $U$.

For widths from equation (26) we have

$$\Gamma_r(E) = -\frac{S}{S + 2Q_r}\Gamma_r^0 \sum_{k=1}^{M} \rho_k(E) \left|g_{rk}\right|^2. \tag{29}$$

Formulas (25)-(29) provide method's algorithm. For any particular situation, $N = 2$, $N = 3$, etc., coefficients $S$, $Q_r$, $F_{qr}$, $G_{qr}$ are given by very simple expressions presented in Appendix.

In case of *one resonan*ce this approach reduces to the traditional BW function. Matrix $S(E)$ is

$$S_{ij} = \delta_{ij} + 2i\sqrt{\rho_i \rho_j}\,\frac{\Gamma_1^0 g_{1i} g_{1j} / 2}{m_1 - E - i\Gamma_1(E)/2}, \qquad i, j = 1, \ldots, M. \tag{30}$$

Unitarity leads to $g_{1i}^y \equiv 0,\ (i = 1, \ldots, M)$, i.e. $\vec{g}_1 = \vec{g}_1^x$ is real vector and $\Gamma_1^0 g_{1i} g_{1j} = \sqrt{\Gamma_{ri}\Gamma_{rj}}$. With (29) we obtain the analogue of Flatte's formula:

$$T_{ij} = \frac{\Gamma_1^0 g_{i1} g_{1j} / 2}{m_1 - E - \frac{i}{2}\Gamma_1^0 \sum_{k=1}^{M} \rho_k(E)\left(g_{1k}^x\right)^2}, \tag{31}$$

or in variable *s*

$$T_{ij} = \frac{m_1 \Gamma_1^0 g_{i1} g_{1j}}{m_1^2 - s - i m_1 \Gamma_1^0 \sum_{k=1}^{M} \rho_k(s)\left(g_{1k}^x\right)^2}. \tag{32}$$

In a fitting procedure mass $m_1$ and *M* components of vector $\vec{g}_1 = \vec{g}_1^x$ can be taken as free parameters.

If a resonance is above all the thresholds, its mass is just $m_1$ and the width is

$$\Gamma_1(E) = \Gamma_1^0 \sum_{k=1}^{M} \rho_k(E)\left(g_{1k}^x\right)^2. \tag{33}$$

If a resonance is between thresholds *L* and $(L+1)$, there are two options: to set $\rho_k(E) = 0$ in the energy region below the corresponding threshold, $E < E_k$, or to continue it as $\rho_k => i\sigma_k \left|\rho_k\right|$ ($\sigma_k$ can take the values $\pm 1$ if to consider different Riemann sheets). Then effective mass is

$$\tilde{m}_1(E) = m_1 + \frac{\Gamma_1^0}{2} \sum_{k=L+1}^{M} \sigma_k \left|\rho_k(E)\right|\left(g_{1k}^x\right)^2 \tag{34}$$

and the width is

$$\Gamma_1(E) = \Gamma_1^0 \sum_{k=1}^{L} \rho_k(E)\left(g_{1k}^x\right)^2. \tag{35}$$

In case of *two resonances and one channel*, the unitary BW approach gives the same expression for the amplitude *T* as the *K*-method, as discussed in section 1.

Let us illustrate the method for *two resonances and two channels,* $N = 2$, $M = 2$. The entire problem consists of finding vectors $\vec{g}_r$. For two channels, matrix *U* is

$$U = \begin{pmatrix} 0 & -\alpha \\ \alpha & 0 \end{pmatrix} \tag{36}$$

($\alpha$ is a real parameter, $0 \le \alpha < 1$), i.e.

$$\vec{g}_1^{\,y} = -\alpha \vec{g}_2^{\,x}, \quad \vec{g}_2^{\,y} = \alpha \vec{g}_1^{\,x}. \tag{37}$$

The coefficients in the unitary constraints (26)-(28) are:

$$S = 1 - \alpha^2, \; Q_1 = Q_2 = \alpha^2, \; F_{12} = -\alpha, \; G_{12} = 0. \tag{38}$$

Altogether there is 13 quantities, $m_1$, $m_2$, $\Gamma_1$, $\Gamma_2$, $\vec{g}_1$, $\vec{g}_2$, $\alpha$, related by four equations (37) and three equations (26)-(28). Thus, 6 real quantities can be taken as free (independent) parameters in a potential fitting procedure. It can be, like we do in section 3, masses $m_1$, $m_2$, parameter $\alpha$ and three components of vectors $\vec{g}_r^x$, for instance $g_{11}^x$, $g_{12}^x$ and $g_{22}^x$. The remaining component $g_{21}^x$ is determined from (27-28). When both resonances are above the 2nd threshold, $E_2 < m_1 < m_2$,

$$g_{21}^x = \frac{1}{\rho_1(m_2) g_{11}^x}\left[\frac{2\alpha}{\Gamma_1^0\left(1-\alpha^2\right)^2}(m_1 - m_2) - \rho_2(m_2) g_{12}^x g_{22}^x\right], \tag{39}$$

then components $\vec{g}_r^y$ are obtained with (37). It completes the construction of the amplitudes

$$T_{ij} = \frac{\Gamma_1^0 g_{1i} g_{1j}/2}{m_1 - E - i\Gamma_1(E)/2} + \frac{\Gamma_2^0 g_{2i} g_{2j}/2}{m_2 - E - i\Gamma_2(E)/2}, \tag{40}$$

where the widths $\Gamma_r(E)$ are given by expression (29), and their values are

$$\Gamma_r(m_r) = \frac{1-\alpha^2}{1+\alpha^2}\Gamma_r^0 \sum_{k=1}^{2} \rho_k(m_r)\left|g_{rk}\right|^2. \tag{41}$$

For $m_1 < E_2 < m_2$, the mass of the 1st resonance, $\tilde{m}_1$, is determined by zero of the real part of the denominator of the first term in (40), i.e. by the root of equation

$$E - \tilde{m}_1 - \frac{1-\alpha^2}{2(1+\alpha^2)}\Gamma_1^0 \left|\rho_2(E)\right| \cdot \left|g_{12}\right|^2 = 0; \tag{42}$$

the width is

$$\tilde{\Gamma}_1(\tilde{m}_1) = \frac{1-\alpha^2}{1+\alpha^2}\Gamma_1^0 \rho_2(\tilde{m}_1)\left|g_{12}\right|^2. \tag{43}$$

The component $g_{21}^x$ can be found from (39) with $\tilde{m}_1$ instead of $m_1$.

## 3. Examples: K-matrix vs. unitary BW

Let us consider two resonances in two channels, $S = I + 2iF = I + 2i\sqrt{\rho}T\sqrt{\rho}$, $\rho$ is a diagonal matrix $2\times 2$. In the unitary BW expression

$$S_{ij} = \delta_{ij} + 2i\sqrt{\rho_i \rho_j}\sum_{r=1}^{2}\frac{m_r \Gamma_r(m_r) g_{ri} g_{rj}}{m_r^2 - s - i m_r \Gamma_r(s)} \tag{44}$$

there are 6 independent parameters (one of the ways to choose them is shown in the previous section). The *K*-matrix pole terms in expression (10),

$$K_{ij} = \gamma_{1i}\gamma_{1j}\frac{m_1\Gamma_1^0}{m_1^2 - s} + \gamma_{2i}\gamma_{2j}\frac{m_2\Gamma_2^0}{m_2^2 - s}, \tag{45}$$

contain the same number of independent parameters: $m_r$, $\Gamma_r^0$ and two $\gamma_{ri}$ (with $\gamma$'s normalized, $\sum_i \gamma_{ri}^2 = 1$). The amplitudes are

$$F = \frac{\sqrt{\rho} K \sqrt{\rho}}{1 - i\rho K} = \frac{1}{1 - \rho_1\rho_2 D - i\left(\rho_1 K_{11} + \rho_2 K_{22}\right)} \begin{pmatrix} \rho_1\left(K_{11} - i\rho_2 D\right) & \sqrt{\rho_1\rho_2} K_{12} \\ \sqrt{\rho_1\rho_2} K_{21} & \rho_2\left(K_{22} - i\rho_1 D\right) \end{pmatrix}, \qquad (46)$$

where $D = K_{11}K_{22} - K_{12}^2$.

Let us start with the statement that *K*-matrix amplitudes have zero, or very close to zero, values between resonance locations. This feature directly follows from expression (7) and therefore remains in any modification of the *K*-matrix method. In case of two resonances, the zero of equation $T_{12}(E) = 0$ (i.e. $K_{12}(E) = 0$) is located between the poles $m_{1,2}$ and given by a linear equation; in case of three states this is a quadratic equation – this result immediately follows from expressions (10) and (46). The zeros in $T_{11}$ and $T_{22}$ are slightly shifted from that location. If relatively narrow resonances are far from each other and background is neglected, the amplitudes approximately presented by a sum of isolated BW functions really become zero between the peaks. But in a situation when the states physically overlap, this feature of the *K*-matrix method can be considered as a defect - the resonances in *K*-matrix scattering amplitudes are always isolated and *actually do not overlap*. Same feature translates on parametrizations that exploit the *K*-matrix for production amplitudes, $A_p = \sum_i (I - i\rho K)_{pi}^{-1} P_i$ (vector $P_i$ has the same poles as *K*, it can also contain a polynomial). Nonzero reference level in mass projection of a Dalitz plot can only conceal this feature. This phenomena can be checked with the interactive software prepared as supplemental material to this paper (it can be sent by request).

It is instructive to present simple examples demonstrating the features of the methods rather than considering actual physical problems in which the demonstration of the approach, which is the goal of this paper, could be hidden in the details of complexity and ambiguity of situation.

Let us first consider $\sim 0.15 \div 0.3\,\mathrm{GeV}$ wide *overlapping* states (peaks) around 1.3 and $1.6\,\mathrm{GeV}$ in amplitudes $\left|F_{ij}(E)\right|^2$, also consider different locations of the 2nd channel threshold: below both states – at $E_2 = 1.22$, and between the states – at $E_2 = 1.38$ (1st threshold is fixed at $E_1 = 0.5$, all quantities are in GeV). Then we generate the "data" which qualitatively correspond to these situations. Technically, we draw smooth curves (different peak heights and widths can be considered), discretize and randomize these curves, introduce dispersions (error bars at each point are generated randomly with the upper limit of 0.05). Two examples of such data are presented in Fig. 2 for $E_2 = 1.22$ and Fig. 3 for $E_2 = 1.38$. The peaks overlap and, at least in one channel, not well resolved, like it often happens in real situations. Then we fit these data sets with the BW formulas (44) – blue lines, and *K*-matrix (46) – red lines.

In calculations we use the phase-factors $\rho_i(s) = \sqrt{(s - s_i)/s}$ (they can be rewritten by including the barrier factors – here we want to avoid unnecessary details to make the subject more clear). In $T_{ij}$, continuation $\rho_2(s) => \pm i|\rho_2(s)|$ below $E_2$ can be used.

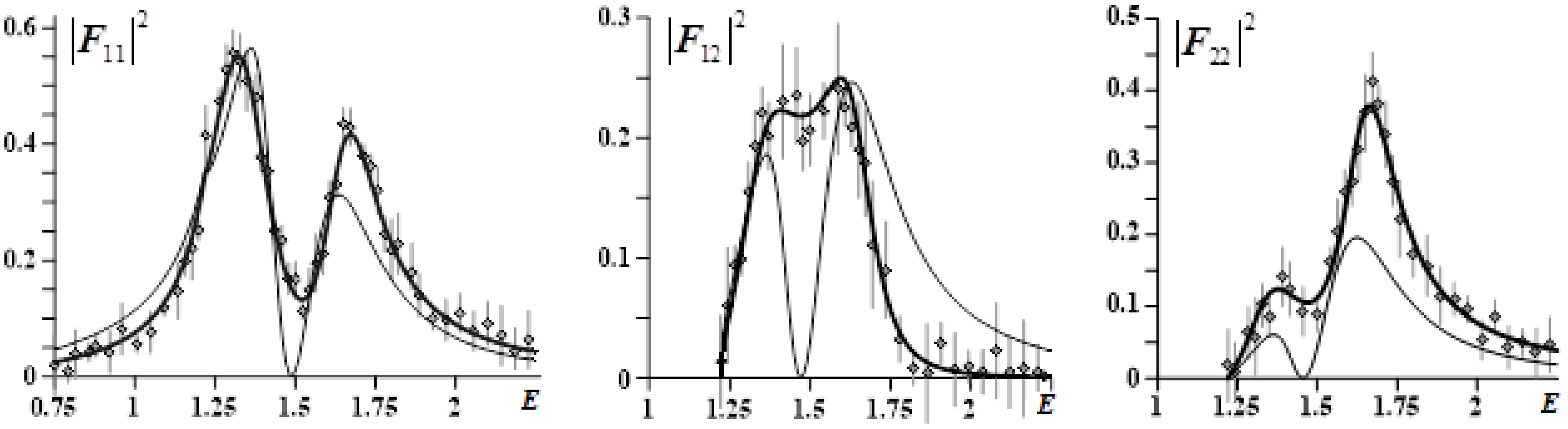

Fig.2. Thin lines – *K*-matrix (46), thick lines– plot BW formulas (44); threshold $E_2 = 1.22$.

In formulas (44) for $E_2 = 1.22$ the values of six independent parameters are:
$m_1 = 1.35, \quad m_2 = 1.65, \quad \alpha = 0.07, \quad g_{11}^x = -0.43, \quad g_{12}^x = 0.41, \quad g_{22}^x = 0.49$,
resulting in $\chi^2 / d = 0.49$ ($d$ is degrees of freedom number);
for $E_2 = 1.38$: $m_1 = 1.36, \quad m_2 = 1.65, \quad \alpha = 0.08, \quad g_{11}^x = -0.42, \quad g_{12}^x = 0.38, \quad g_{22}^x = 0.36$,
resulting in $\chi^2 / d = 0.53$.
The remaining parameters can be calculated using these six independent ones. (After the $g_{ri}$ values are found, parameter $\alpha$, which is a technical one, is not needed any longer.)

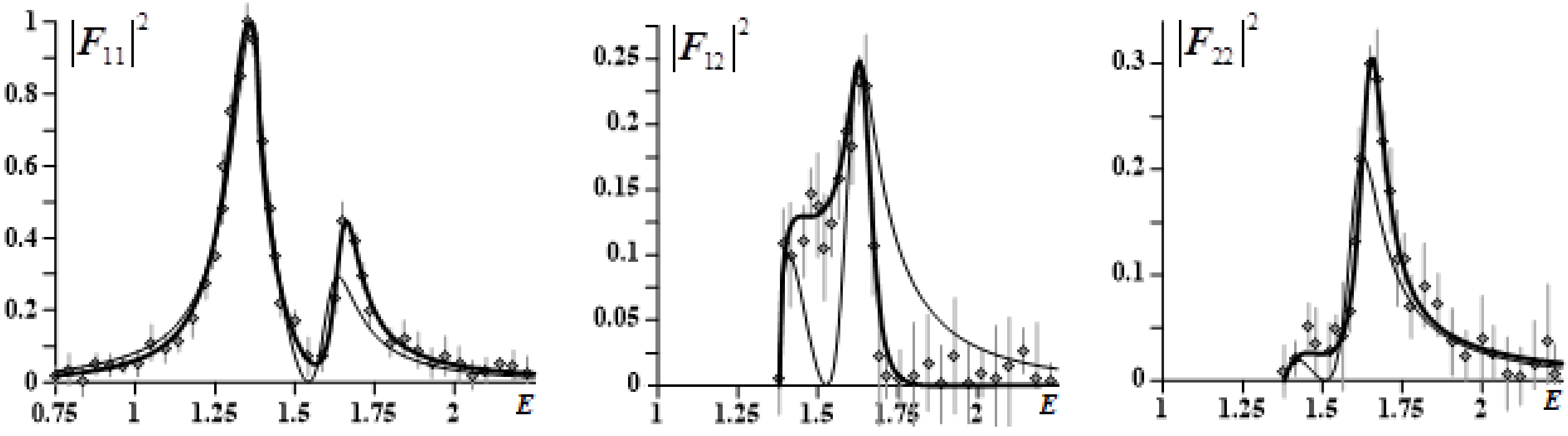

Fig.3. Thin lines – *K*-matrix (46), thick lines– plot BW formulas (44); threshold $E_2 = 1.38$.

In Table 1 we collect the results in the BW approach (branching ratios $B_{ri}$ are calculated with formulas (21), phases $\varphi_{ij}^{r}$ can be obtained from the $g_{ri}$ values). By direct substitution of these $g_{ri}$ in (44) it can be checked that matrix $S(E)$ is unitary for any *E*.

**TABLE 1.** BW parameters (6 independent) for two threshold positions

| threshold $E_2=1.22$ | | | | | | threshold $E_2=1.38$ | | | | | |
|---|---|---|---|---|---|---|---|---|---|---|---|
| $m_r$ | $\Gamma_r$ | $B_{r1}$ | $B_{r2}$ | $g_{r1}$ | $g_{r2}$ | $m_r$ | $\Gamma_r$ | $B_{r1}$ | $B_{r2}$ | $g_{r1}$ | $g_{r2}$ |
| 1.32 | 0.26 | 53 | 47 | –0.43–*i*0.03 | 0.41–*i*0.04 | 1.36 | 0.19 | 55 | 45 | –0.43–*i*0.02 | –0.38–*i*0.03 |
| 1.65 | 0.318 | 44 | 56 | 0.43–*i*0.03 | 0.49+*i*0.03 | 1.65 | 0.14 | 42 | 58 | 0.30–*i*0.03 | 0.36–*i*0.03 |

In K-matrix formulas (46) the values of independent parameters for $E_2=1.22$ are found to be:
$m_1=1.36,\ m_2=1.63,\ \Gamma_1=0.27,\ \Gamma_2=0.37,\ \gamma_{11}=0.77,\ \gamma_{21}=0.68$,
resulting in $\chi^2/d=8.72$;
for $E_2=1.38$: $m_1=1.37,\ m_2=1.63,\ \Gamma_1=0.32,\ \Gamma_2=0.19,\ \gamma_{11}=0.73,\ \gamma_{21}=0.63$,
resulting in $\chi^2/d=4.75$.

Table 2 contains the K-matrix parameters. Branching ratios are defined as $B_{ri}=\gamma_{ri}^2$.

**TABLE 2.** K-matrix parameters (6 independent) for two threshold positions

| threshold $E_2=1.22$ | | | | threshold $E_2=1.38$ | | | |
|---|---|---|---|---|---|---|---|
| $m_r$ | $\Gamma_r$ | $B_{r1}$ | $B_{r1}$ | $m_r$ | $\Gamma_r$ | $B_{r1}$ | $B_{r2}$ |
| 1.36 | 0.27 | 59 | 41 | 1.37 | 0.32 | 54 | 46 |
| 1.63 | 0.37 | 47 | 53 | 1.63 | 0.20 | 40 | 60 |

Large values of $\chi^2$ only reflect the fact that the *K*-matrix method inadequately describes the regions between the peaks when *all the amplitudes* (cross-sections) are not close to zero in these locations. But if the amplitudes have not only the poles, but also the zeros, in other words, when resonances are well resolved and do not overlap, both the *K*-matrix and BW descriptions lead to close results (obviously, a background - any additional plateau in data in resonant area - will lead to a discrepancy). This situation is presented in Fig. 4 in which the data have substantial dips between the resonances in all three channels (for brevity we consider only one threshold location $E_2=1.22$). The quality of fits is practically the same - in each method $\chi^2/d\approx0.5$. The resonance parameters for both methods are collected in Table 3.

**TABLE 3.** BW and K-matrix parameters (6 independent) for data in Fig. 4.

| BW | | | | | | K | | | |
|---|---|---|---|---|---|---|---|---|---|
| $m_r$ | $\Gamma_r$ | $B_{r1}$ | $B_{r2}$ | $g_{r1}$ | $g_{r2}$ | $m_r$ | $\Gamma_r$ | $B_{r1}$ | $B_{r2}$ |
| 1.30 | 0.19 | 49 | 51 | –0.39–*i*0.02 | –0.37–*i*0.03 | 1.30 | 0.28 | 45 | 55 |
| 1.60 | 0.09 | 18 | 82 | –0.11–*i*0.05 | –0.25+*i*0.06 | 1.60 | 0.12 | 27 | 73 |

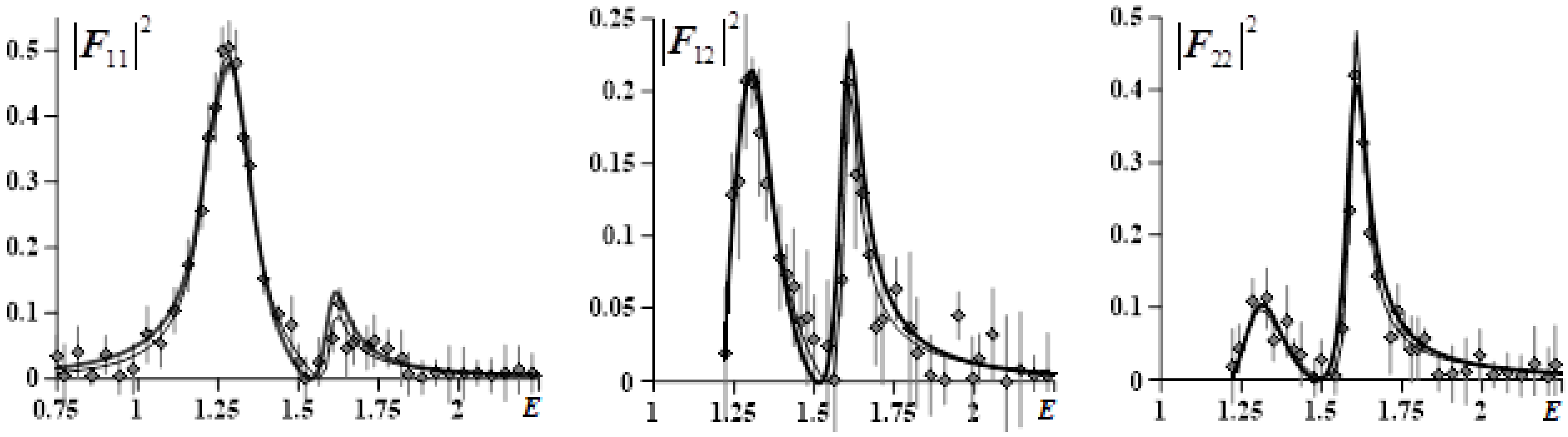

Fig. 4. Thin lines – *K*-matrix (46), thick lines– BW formulas (44); threshold $E_2 = 1.22$.

In the end, we want to highlight the simplicity of the unitary BW method: finding $g_{ri}$ with formulas (25)-(28) is an easy straightforward procedure and with these vectors we have the final expressions for partial amplitudes.

## 4. Background in Breit-Wigner unitary method

In practice, resonances are accompanied by background. With polynomial in $K_{ij}$ the dips in amplitudes between the poles remain (this also can be checked with the accompanying software; an obvious restriction on polynomial coefficients is that resonances manifestation should retain). Also notice that polynomial terms in (3) do not allow to present background in the quantum mechanics form, $S^B_{ij} = e^{i(\beta_i + \beta_j)}$ in which even the number of parameters is different, for instance for two channels and energy-independent background two parameters $\beta_{1,2}$ versus three $a_{ij}$ in $K_{ij}$.

In the BW scheme $S$ -matrix can be presented as a sum of resonant and background terms [8]

$$S_{ij} = B_{ij} + 2i\sqrt{\rho_i \rho_j} \sum_{r=1}^{N} \frac{m_r \Gamma_r^0 A_{ri} A_{rj}}{m_r^2 - s - i m_r \Gamma_r(s)}. \quad (47)$$

Background matrix *B* should be unitary and symmetric (*T*-invariant) as well as matrix *S*.

If matrix *B* is diagonal,

$$B_{ij} = \delta_{ij} e^{i\beta_i} \qquad (i, j = 1,...,M),$$

then vectors $\vec{A}_r$ can be written as

$$A_{rk} = e^{i\beta_k} g_{rk} = \left(\cos\beta_k + i\sin\beta_k\right)\left(g_{rk}^x + i g_{rk}^y\right),$$

where $g_{ri}$ are determined by the procedure described in previous sections. Matrix *S* in such a situation is

$$S_{ij} = e^{i(\beta_i+\beta_j)}\left[\delta_{ij} + 2i\sqrt{\rho_i\rho_j}\sum_{r=1}^{N} e^{i(\varphi_{ri}+\varphi_{rj})}\frac{m_r\Gamma_r^0 |g_{ri}|\cdot|g_{rj}|}{m_r^2 - s - i m_r\Gamma_r}\right]. \tag{48}$$

Note that diagonal part of background given by the exponents $e^{i\beta_k}$ does not contribute to $|F_{ij}|^2$ for $i \neq j$.

If matrix *B* is not diagonal, i.e. background potentially is not zero in all channels, it is helpful to present it in the form [3], $B = We^{2i\vec{\beta}}W^T$, where *W* is real orthogonal matrix (orthogonal matrix is a real square matrix such that $A^{-1} = A^T$); $e^{2i\vec{\beta}}$ is a diagonal matrix. In case when matrix *B* is diagonal, $W = I$. Writing down matrix *B* as $B = bb^T$ with $b = We^{i\vec{\beta}}$, we can define $\vec{g}_r = b^T\vec{A}_r$. Then $\vec{A}_r = b\vec{g}_r$ and matrix *S* can be written as $S = b\tilde{S}b^T$, where $\tilde{S}$ does not contain background:

$$\tilde{S}_{ij} = \delta_{ij} + 2i\sqrt{\rho_i\rho_j}\sum_{r=1}^{N}\frac{m_r\Gamma_r^0 g_{ri} g_{rj}}{m_r^2 - s - i m_r\Gamma_r}. \tag{49}$$

Obviously, $S$ is unitary if matrix $\tilde{S}$ is unitary. Therefore, we can independently determine vectors $g_{ri}$ in $\tilde{S}$ and then return to matrix $S$.

Matrix *W* can be constructed as a chain of $M(M-1)/2$ rotations:

$$W = R_1 \cdot R_2 \cdot ... \cdot R_{M(M-1)/2}. \tag{50}$$

Matrix $R_k = \left\{r_{ij}^{(k)}\right\}$ $(i, j = 1,...,M)$ is naturally to take in the form:

$$r_{pp}^{(k)} = r_{qq}^{(k)} = \cos\psi_k, \quad r_{pq}^{(k)} = -r_{qp}^{(k)} = \sin\psi_k, \tag{51}$$

$p = 1,...,M,$ $q = p+1,...,M$ ; $\psi_k \in [0, 2\pi)$ are rotation angles.

For instance, for three channels

$$b = W\begin{pmatrix} e^{i\beta_1} & 0 & 0 \\ 0 & e^{i\beta_2} & 0 \\ 0 & 0 & e^{i\beta_3} \end{pmatrix},$$

where

$$W = R_1 R_2 R_3 = \begin{pmatrix} \cos\psi_1 & \sin\psi_1 & 0 \\ -\sin\psi_1 & \cos\psi_1 & 0 \\ 0 & 0 & 1 \end{pmatrix} \begin{pmatrix} 1 & 0 & 0 \\ 0 & \cos\psi_2 & \sin\psi_2 \\ 0 & -\sin\psi_2 & \cos\psi_2 \end{pmatrix} \begin{pmatrix} \cos\psi_3 & 0 & \sin\psi_3 \\ 0 & 1 & 0 \\ -\sin\psi_3 & 0 & \cos\psi_3 \end{pmatrix}.$$

The quantities $\beta_k$, $\psi_k$ can be considered as fit parameters (they can be functions of energy).

Then, we have

$$F_{ij} = (S_j - I)/2i = (B_F)_{ij} + \sqrt{\rho_i \rho_j} \sum_{r=1}^{N} \frac{m_r \Gamma_r^0 A_{ri} A_{rj}}{m_r^2 - s - i m_r \Gamma_r(s)}, \tag{52}$$

where matrix

$$B_F = W \begin{pmatrix} \sin\beta_1 e^{i\beta_1} & \dots & 0 \\ \dots & \dots & \dots \\ 0 & \dots & \sin\beta_M e^{i\beta_M} \end{pmatrix} W^T. \tag{53}$$

Thus, background is very easy to implement into unitary BW scheme simply using formulas (52) and (53).

To demonstrate the scheme and its convenience, let us consider an example with two resonances and three channels; parameters are presented in Table 4. Then the background is added to these unitary BW amplitudes. In Fig. 5 graphs of $|F_{11}(E)|^2$ and $|F_{12}(E)|^2$ (chosen as examples) are shown for background taken in form of diagonal and non-diagonal matrix. For shortness, we consider energy-independent background.

Table 4. BW pole functions

| $r$ | $m_r$ | $\Gamma_r$ | $B_{r1}$ (%) | $B_{r2}$ (%) | $B_{r3}$ (%) |
|---|---|---|---|---|---|
| Resonance 1 | 1.30 | 0.10 | 40 | 40 | 20 |
| Resonance 2 | 1.70 | 0.30 | 20.8 | 59.4 | 19.8 |

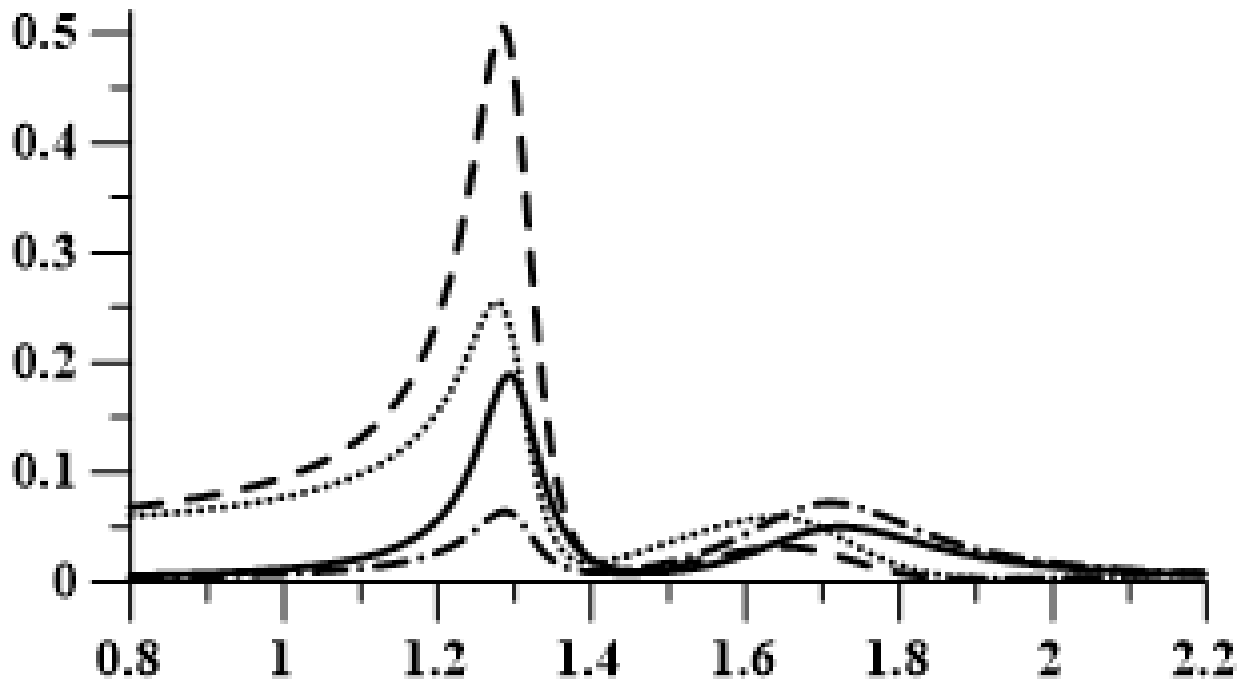

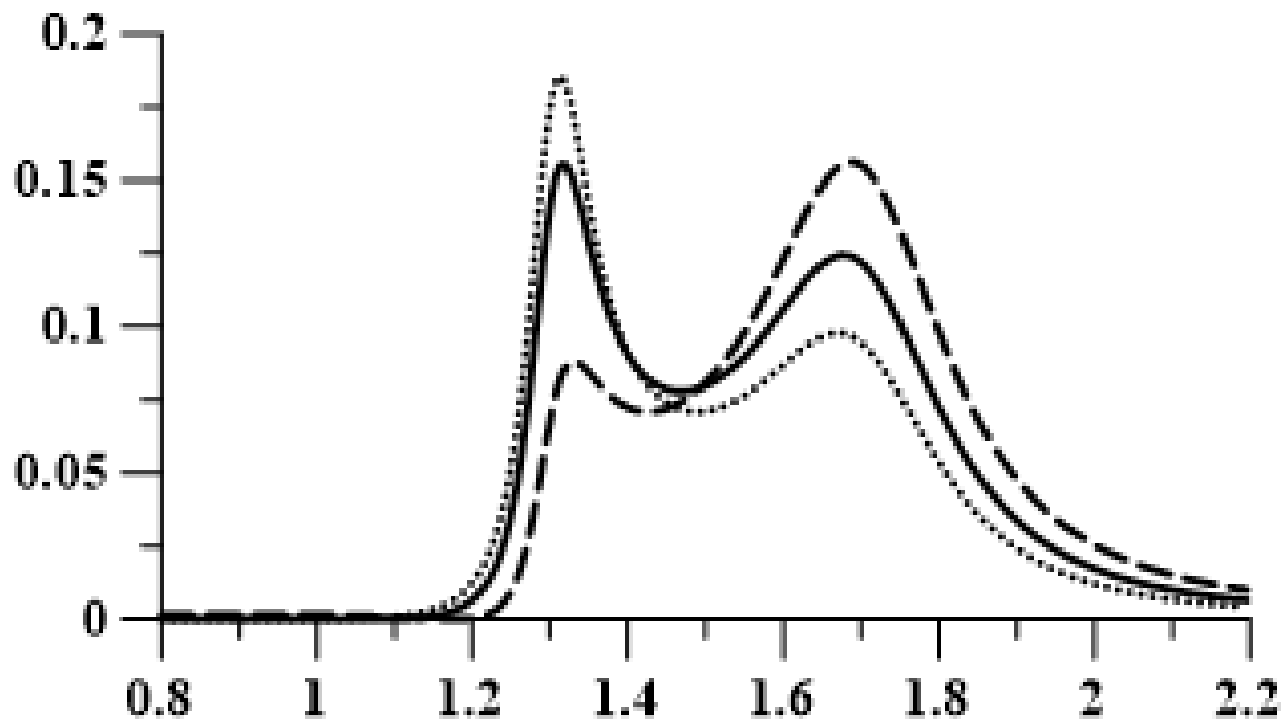

Fig. 5. Left: $|F_{11}(E)|^2$. Solid line –no background, $\beta_k = 0$, $\psi_k = 0$ $(k = 1,2,3)$; dotted line – $\beta_k = \pi/18$, $\psi_k = 0$; dashed line – $\beta_k = \pi/18$, $\psi_k = \pi/18$; dashed-dotted line – $\beta_k = 0$, $\psi_k = -\pi/18$. Right: $|F_{12}(E)|^2$. $\beta_k = 0$ for all curves: solid line – $\psi_k = 0$ (no background); dotted line – $\psi_k = \pi/18$; dashed line – $\psi_k = -\pi/18$.

It is seen, that even a $10^o$ *diagonal* background modifies the resonance shapes. Background obtained with *rotation* affects even peak locations. For example, second peak in $|F_{11}(E)|^2$ is shifted by about 0.15 from $m_2$, the "correct" one.

In the *K*-matrix method, besides that there is no regular way to incorporate a background, the amplitudes like in Figs. 5, cannot be obtained even technically by adding polynomial terms to $K_{ij}$ - the zeros in the amplitudes do not go away and the plateau in the resonant area does not appear (this also can be investigated with the provided software).

Because apriori there is no criteria to select background parametrization type, this creates an uncertainty in masses, widths and branching ratios. If one particular parametrization fits the data substantially better than others, it gives the practical criteria. If this is not the case, a simplified approach may be the following: first to find the background by smoothing the data [9,10] which removes all the peaks, then to subtract so obtained background curves from the data and to analyze the remaining data with the BW unitarization scheme without background – this approach was used in work [9] to analyze overlapping $\omega'(1400)$ and $\omega'(1600)$ resonances in approximation of constant widths.

## 5. Conclusion

Obviously, the interference between resonances is the central part of analysis and interpretation. The unitary BW approach is conceptually very simple and provides parametrization in terms which have explicit physical meaning, a background is easy to include in the scheme in the standard form through the phase shifts. The amplitudes very naturally contain quantum mechanics interference phases. The production channels in the BW scheme do not need a special treatment, corresponding $g_{rp}$ (*p* denotes these channels) having different physical nature, just may have a different order of magnitude comparing to other $g_{ri}$.

It is important to be aware that actually overlapping (not resolved) resonances cannot be adequately described within the *K*-matrix parametrizations. Another problem is that the *K*-matrix amplitudes cannot be separated in resonant and background contributions.

From theoretical point of view it worth knowing that the problem formulated long time ago – to present the unitary *S*-matrix as a sum of BW terms – has a simple and regular solution. Papers on overlapping resonances analysis often begin with the statement that a sum of BW functions violates the unitarity and alternative methods should be used. With this paper we want to attract attention to the fact that the unitary BW representation for several states is possible. It can be used in fitting procedures for the same problems as the standard *K*-matrix method which provides a convenient parametrization technique widely used in analyses. Despite the *K*-matrix approach serves perfectly fine as unitary parametrization, the terms of its parameters, "masses", "widths", "branching ratios", are only borrowed from the BW description through the comparison with a single resonance case. Developing another unitarity preserving, model-independent method to describe overlapping resonances, which parameters have direct physical meaning, can be a good addition or alternative to the *K*-matrix method.

## Appendix

## Construction of unitary *S* matrix

For *N* resonances and *M* channels the *S* matrix is given by (background contribution is considered in section 4) expressions

$$S(E) = I + 2i\sqrt{\rho}T\sqrt{\rho} \qquad \text{(A1)}$$

with (variable *E* is used only to simplify formulas)

$$T(E) = \sum_{r=1}^{N} \frac{\vec{g}_r \vec{g}_r / 2}{\varepsilon_r(E) - E}. \qquad \text{(A2)}$$

The unitarity condition $S^{\dagger}(E)S(E) \equiv I$ gives

$$T - T^{+} = 2iT\rho T^{+},$$

or

$$i\sum_{r=1}^{N}\left[\frac{g_{ri}g_{rj}}{\varepsilon_r - E} - \frac{g_{ri}^{*}g_{rj}^{*}}{\varepsilon_r^{*} - E}\right] + \sum_{l=1}^{N}\sum_{r=1}^{N} V_{lr}\frac{g_{ri}g_{lj}^{*}}{(\varepsilon_r - E)(\varepsilon_l^{*} - E)} \equiv 0, \qquad \text{(A3)}$$

where

$$V_{lr} = \sqrt{\Gamma_l^0 \Gamma_r^0}\sum_{k=1}^{M} \rho_k(E) g_{lk}^{*} g_{rk}. \qquad \text{(A4)}$$

Multiplying (A3) by the product $\prod_{k=1}^{N}(\varepsilon_k - E)(\varepsilon_k^{*} - E)$ we obtain

$$\begin{aligned}&\sum_{r=1}^{N} B_r\left[-ig_{ri}^{*}g_{rj}^{*}(\varepsilon_r - E) + ig_{ri}g_{rj}(\varepsilon_r^{*} - E) + V_{rr}g_{ri}g_{rj}^{*}\right] + \\ &+\sum_{r=1}^{N-1}\sum_{l=r+1}^{N} B_{rl}\left[V_{lr}g_{ri}g_{lj}^{*}(\varepsilon_l - E)(\varepsilon_r^{*} - E) + V_{rl}g_{li}g_{rj}^{*}(\varepsilon_l^{*} - E)(\varepsilon_r - E)\right] \equiv 0,\end{aligned} \qquad \text{(A5)}$$

with $B_r \equiv \prod_{k\neq r}^{N}(E - \varepsilon_k)(E - \varepsilon_k^{*})$, $B_{rl} \equiv \prod_{k\neq r,l}^{N}(E - \varepsilon_k)(E - \varepsilon_k^{*})$.

To satisfy the unitarity relation (A3), it is necessary and sufficient that coefficients of the polynomial (A5) be zero for all powers of $E^k$. The coefficient of $E^k$ is

$$\begin{aligned}&\sum_{r=1}^{N}\left\{B_r^{(k)}V_{rr}g_{ri}g_{rj}^{*} + 2B_r^{(k-1)}\mathrm{Im}(g_{ri}g_{rj}) + 2B_r^{(k)}\left[-\varepsilon_r^{x}\,\mathrm{Im}(g_{ri}g_{rj}) + \varepsilon_r^{y}\,\mathrm{Re}(g_{ri}g_{rj})\right]\right\} + \\ &+\sum_{r=1}^{N-1}\sum_{l=r+1}^{N}\left\{B_{rl}^{(k-2)}\left[V_{lr}g_{ri}g_{lj}^{*} + V_{lr}^{*}g_{li}g_{rj}^{*}\right] - B_{rl}^{(k-1)}\left[V_{lr}g_{ri}g_{lj}^{*}(\varepsilon_l + \varepsilon_r^{*}) + \right.\right. \\ &\left.\left. +V_{lr}^{*}g_{li}g_{rj}^{*}(\varepsilon_l^{*} + \varepsilon_r)\right] + B_{rl}^{(k)}\left[V_{lr}g_{ri}g_{lj}^{*}\varepsilon_l\varepsilon_r^{*} + V_{lr}^{*}g_{li}g_{rj}^{*}\varepsilon_l^{*}\varepsilon_r\right]\right\} = 0.\end{aligned} \qquad \text{(A6)}$$

Here $B_r^{(k)}$ and $B_{rl}^{(k)}$ are coefficients at the powers $E^k$ in the polynomials $B_r$ and $B_{rl}$, respectively.

Equating to zero the coefficient at the highest degree $E^{2N-1}$ with taking into account that $B_r$ is the polynomial of degree $(2N-2)$ and $B_{rl}$ is the polynomial of degree $(2N-4)$, we obtain:

$$\sum_{r=1}^{N} \mathrm{Im}\left(g_{ri} g_{rj}\right) = 0 \quad \text{or} \quad \sum_{r=1}^{N} \left[ g_{ri}^x g_{rj}^y + g_{ri}^y g_{rj}^x \right] = 0, \tag{A7}$$

where $i, j = 1, ..., M$, $\vec{g}_r = \vec{g}_r^{\,x} + i\vec{g}_r^{\,y}$.

The key moment of the method is that we construct the imaginary parts $\vec{g}_r^{\,y}$ of vectors $\vec{g}_r$ as linear combinations of their real parts $\vec{g}_r^{\,x}$ ($r = 1, ..., N$):

$$\vec{g}_r^{\,y} = u_{r1}\vec{g}_1^{\,x} + u_{r2}\vec{g}_2^{\,x} + ... + u_{rN}\vec{g}_N^{\,x}. \tag{A8}$$

Substituting expression (A8) into relation (A7), we obtain $u_{rr} = 0$ and $u_{rk} = -u_{kr}$ thus $U$ is anti-symmetric matrix. Equation (A8) gives $NM$ relations involving $N(N-1)/2$ parameters (matrix elements $u_{ij}$).

Next, equate to zero the coefficient at $E^{2N-2}$:

$$\begin{aligned} &\sum_{r=1}^{N} \left\{ 2B_r^{(2N-3)} \mathrm{Im}(g_{ri} g_{rj}) + B_r^{(2N-2)} \left[ V_{rr} g_{ri} g_{rj}^* - 2\varepsilon_r^x \mathrm{Im}(g_{ri} g_{rj}) + 2\varepsilon_r^y \mathrm{Re}(g_{ri} g_{rj}) \right] \right\} + \\ &+ \sum_{r=1}^{N-1} \sum_{l=r+1}^{N} \left\{ B_{rl}^{(2N-4)} \left[ V_{lr} g_{ri} g_{lj}^* + V_{lr}^* g_{li} g_{rj}^* \right] - B_{rl}^{(2N-3)} \left[ V_{lr} g_{ri} g_{lj}^* (\varepsilon_l + \varepsilon_r^*) + \right. \right. \\ &\left. \left. + V_{lr}^* g_{li} g_{rj}^* (\varepsilon_l^* + \varepsilon_r) \right] + B_{rl}^{(2N-2)} \left[ V_{lr} g_{ri} g_{lj}^* \varepsilon_l \varepsilon_r^* + V_{lr}^* g_{li} g_{rj}^* \varepsilon_l^* \varepsilon_r \right] \right\} = 0. \end{aligned} \tag{A9}$$

Because $B_r^{(2N-3)} = -2\sum_{k \neq r} \varepsilon_k^x$, $B_r^{(2N-2)} = B_{rl}^{(2N-4)} = 1$, $B_{rl}^{(2N-2)} = B_{rl}^{(2N-3)} = 0$, we obtain

$$\begin{aligned} &\sum_{r=1}^{N} \left[ -\sum_{k \neq r} 4\varepsilon_k^x \mathrm{Im}(g_{ri} g_{rj}) + 2\varepsilon_r^y \mathrm{Re}\,(g_{ri} g_{rj}) - 2\varepsilon_r^x \mathrm{Im}\,(g_{ri} g_{rj}) + V_{rr} g_{ri} g_{rj}^* \right] + \\ &\qquad + \sum_{r=1}^{N-1} \sum_{l=r+1}^{N} \left[ V_{lr} g_{ri} g_{lj}^* + V_{lr}^* g_{li} g_{rj}^* \right] = 0. \end{aligned} \tag{A10}$$

Taking into account expression (A8), we can write

$$
\begin{aligned}
g_{ri}g_{rj} &= \left[ g_{ri}^x g_{rj}^x - \sum_{k=1}^{N}\sum_{\nu=1}^{N} u_{rk}u_{r\nu} g_{ri}^x g_{\nu j}^x \right] + i\sum_{k=1}^{N} u_{rk}\left[ g_{ri}^x g_{kj}^x + g_{rj}^x g_{ki}^x \right],\\
g_{ri}g_{rj}^* &= \left[ g_{ri}^x g_{rj}^x + \sum_{k=1}^{N}\sum_{\nu=1}^{N} u_{rk}u_{r\nu} g_{ki}^x g_{\nu j}^x \right] + i\sum_{k=1}^{N} u_{rk}\left[ -g_{ri}^x g_{kj}^x + g_{rj}^x g_{ki}^x \right],\\
g_{ri}g_{lj}^* &= \left[ g_{ri}^x g_{lj}^x + \sum_{k=1}^{N}\sum_{\nu=1}^{N} u_{rk}u_{l\nu} g_{ki}^x g_{\nu j}^x \right] + i\sum_{k=1}^{N}\left[ -u_{lk} g_{ri}^x g_{kj}^x + u_{rk} g_{lj}^x g_{ki}^x \right],\\
g_{li}g_{rj}^* &= \left[ g_{li}^x g_{rj}^x + \sum_{k=1}^{N}\sum_{\nu=1}^{N} u_{lk}u_{r\nu} g_{ki}^x g_{\nu j}^x \right] + i\sum_{k=1}^{N}\left[ -u_{rk} g_{li}^x g_{kj}^x + u_{lk} g_{rj}^x g_{ki}^x \right].
\end{aligned}
\tag{A11}
$$

In order the coefficient in (A10) at $E^{2N-2}$ to be zero, it is necessary to set to zero the coefficients of all products $g_{\mu i}^x g_{\nu j}^x$, $(\mu,\nu = 1,...,N)$. We substitute expressions (A11) into relation (A10) and equate to zero the real and imaginary parts of these coefficients. This yields:

$$
2\varepsilon_\mu^y - 2\sum_{r\neq\mu}\varepsilon_\mu^y u_{r\mu}^2 + V_{\mu\mu} + \sum_{r\neq\mu} V_{rr}u_{r\mu}^2 + 2\sum_{r=1}^{N-1}\sum_{l=r+1}^{N} V_{lr}^x u_{r\mu}u_{l\mu} + 2\sum_{r\neq\mu} V_{r\mu}^y u_{r\mu} = 0, \tag{A12}
$$

$$
\begin{aligned}
&-2\sum_{r=1}^{N}\varepsilon_r^y u_{r\mu}u_{r\nu} + 2u_{\mu\nu}(\varepsilon_\mu^x - \varepsilon_\nu^x) + \sum_{r=1}^{N} V_{rr}u_{r\mu}u_{r\nu} + V_{\mu\nu}^x + \\
&\qquad + \sum_{r=1}^{N-1}\sum_{l=r+1}^{N} V_{lr}^x [u_{r\mu}u_{l\nu} + u_{l\mu}u_{r\nu}] + \sum_{r\neq\mu,\nu}^{N} [V_{r\mu}^y u_{r\nu} + V_{r\nu}^y u_{r\mu}] = 0,
\end{aligned}
\tag{A13}
$$

$$
u_{\nu\mu}(V_{\mu\mu} + V_{\nu\nu}) + V_{\nu\mu}^y + \sum_{r=1}^{N-1}\sum_{l=r+1}^{N} V_{lr}^y [u_{r\mu}u_{l\nu} - u_{l\mu}u_{r\nu}] + \sum_{r\neq\mu,\nu}^{N} [V_{r\nu}^x u_{r\mu} + V_{r\mu}^x u_{r\nu}] = 0. \tag{A14}
$$

Thus, we obtain $N^2$ simultaneous equations linear in the scalar products $V_{rr}$, $V_{lr}^x$ and $V_{lr}^y$: $N$ equations (A12), $N(N-1)/2$ equations (A13), and $N(N-1)/2$ equations (A14). The coefficients at $g_{\mu i}^x g_{\nu j}^x$ and $g_{\nu i}^x g_{\mu j}^x$ are identical because equations (A13) and (A14) are symmetrical under $\mu$ and $\nu$ interchange.

Solving these equations, we obtain (recall that $V_{lr} = \sqrt{\Gamma_l^0\Gamma_r^0}\sum_{k=1}^{M}\rho_k(E) g_{lk}^* g_{rk}$):

$$V_{rr} = \Gamma_r^0 \sum_{k=1}^{M} \rho_k(E)\left[\left(g_{rk}^x\right)^2 + \left(g_{rk}^y\right)^2\right] = -\frac{2}{S}\left[S + 2Q_r\right]\varepsilon_r^y,$$

$$V_{lr}^x = \sqrt{\Gamma_l^0 \Gamma_r^0} \sum_{k=1}^{M} \rho_k(E)\left[g_{lk}^x g_{rk}^x + g_{lk}^y g_{rk}^y\right] = -\frac{2}{S}\left[F_{lr}\left(\varepsilon_l^x - \varepsilon_r^x\right) + G_{lr}\left(\varepsilon_l^y + \varepsilon_r^y\right)\right], \qquad \text{(A15)}$$

$$V_{lr}^y = \sqrt{\Gamma_l^0 \Gamma_r^0} \sum_{k=1}^{M} \rho_k(E)\left[g_{lk}^x g_{rk}^y - g_{lk}^y g_{rk}^x\right] = -\frac{2}{S}\left[G_{lr}\left(\varepsilon_l^x - \varepsilon_r^x\right) - F_{lr}\left(\varepsilon_l^y + \varepsilon_r^y\right)\right],$$

$$r = 1, ..., N; \;\; l = r+1, ..., N,$$

where

$$S = 1 + \sum_{p=1}^{\left[\frac{N}{2}\right]} (-1)^p \sum_{i_1=1}^{N-2p+1} \sum_{i_2>i_1}^{N-2p+2} \cdots \sum_{i_{2p}>i_{2p-1}}^{N} \begin{vmatrix} i_1 & i_2 & \dots & i_{2p} \\ i_1 & i_2 & \dots & i_{2p} \end{vmatrix},$$

$$Q_r = \sum_{p=1}^{\left[\frac{N}{2}\right]} (-1)^p \sum_{i_1=1}^{N-2p+2} \sum_{i_2>i_1}^{N-2p+3} \cdots \sum_{i_{2p-1}>i_{2p-2}}^{N} \begin{vmatrix} r & i_1 & i_2 & \dots & i_{2p-1} \\ r & i_1 & i_2 & \dots & i_{2p-1} \end{vmatrix},$$

$$\left(i_1, ..., i_{2p-1} \neq r\right),$$

$$F_{lr} = u_{lr} + \sum_{p=1}^{\left[\frac{N}{2}\right]-1} (-1)^{p+1} \sum_{i_1=1}^{N-2p+1} \sum_{i_2>i_1}^{N-2p+2} \cdots \sum_{i_{2p}>i_{2p-1}}^{N} \begin{vmatrix} l & i_1 & i_2 & \dots & i_{2p} \\ r & i_1 & i_2 & \dots & i_{2p} \end{vmatrix},$$

$$\left(i_1, ..., i_{2p} \neq r, l\right),$$

$$G_{lr} = \sum_{p=1}^{\left[\frac{N+1}{2}\right]-1} (-1)^{p+1} \sum_{i_1=1}^{N-2p+2} \sum_{i_2>i_1}^{N-2p+3} \cdots \sum_{i_{2p-1}>i_{2p-2}}^{N} \begin{vmatrix} l & i_1 & i_2 & \dots & i_{2p-1} \\ r & i_1 & i_2 & \dots & i_{2p-1} \end{vmatrix},$$

$$\left(i_1, ..., i_{2p-1} \neq r, l\right).$$

Here

$$\begin{pmatrix} l & i_1 & i_2 & \dots & i_k \\ r & i_1 & i_2 & \dots & i_k \end{pmatrix} = \begin{vmatrix} u_{lr} & u_{li_1} & \dots & u_{li_k} \\ u_{i_1 r} & u_{i_1 i_1} & \dots & u_{i_1 i_k} \\ \dots & \dots & \dots & \dots \\ u_{i_k r} & u_{i_k i_1} & \dots & u_{i_k i_k} \end{vmatrix}$$

is the minor with the rows $l, i_1, i_2, ..., i_k$ and the columns $r, i_1, i_2, ..., i_k$ of matrix $\boldsymbol{U}$ $(i_1, ..., i_{2p} \neq r, l)$. The notations like $[N/2]$ mean an integer part of the expression in the brackets.

Relations (A8) and the scalar products (A15) completely define the conditions imposed on vectors $\vec{g}_r$ $(r=1,...,N)$. It can be shown that if vectors $\vec{g}_r$ satisfy the relations (A8) and (A15), then the coefficients at lower degrees of polynomial (A5) are identically equal zero.

For practical purposes, it is convenient to express the scalar products (A15) in terms of real vectors $\vec{g}_r^{\,x}$:

$$\Gamma_r^0\sum_{k=1}^{M}\rho_k\left(g_{rk}^x\right)^2=\frac{2}{S^2}\left\{-(S+Q_r)^2\varepsilon_r^y+\sum_{i\neq r}\left[2F_{ri}G_{ri}\varepsilon_i^x+\left(F_{ri}^2-G_{ri}^2\right)\varepsilon_i^y\right]\right\}$$

$$\begin{aligned}\sqrt{\Gamma_l^0\Gamma_r^0}\sum_{k=1}^{M}\rho_k g_{lk}^x g_{rk}^x=\frac{2}{S^2}\Big\{&F_{lr}\left[\left(S+Q_r\right)\varepsilon_r^x-\left(S+Q_l\right)\varepsilon_l^x\right]-\\&-G_{lr}\left[\left(S+Q_r\right)\varepsilon_r^y+\left(S+Q_l\right)\varepsilon_l^y\right]+\\&+\sum_{i\neq l,r}\left[\left(F_{li}G_{ri}+F_{ri}G_{li}\right)\varepsilon_i^x+\left(F_{li}F_{ri}-G_{li}G_{ri}\right)\varepsilon_i^y\right]\Big\},\\&r=1,...,N;\;\; l=r+1,...,N.\end{aligned}\qquad\text{(A16)}$$

The domain of the elements of the matrix $U$ is defined by the conditions

$$0\leq\frac{F_{lr}^2+G_{lr}^2}{\left(S+2Q_r\right)\left(S+2Q_l\right)}\leq\frac{\Gamma_r\Gamma_l}{4\left(E_r-E_l\right)^2+\left(\Gamma_r+\Gamma_l\right)^2}\leq 1,$$

which follows from Cauchy-Schwarz inequality $\left(\vec{a}^*,\vec{b}\right)^2\leq\left(\vec{a}^*,\vec{a}\right)\left(\vec{b}^*,\vec{b}\right)$.

It is seen from these inequalities that if resonances are very far from each other, then $u_{rl}\to 0$ and vectors $\vec{g}_r$ become real and orthogonal. In this situation the lengths of vectors $\vec{g}_r$ are equal to the resonance widths $\Gamma_r$, but in general these lengths are larger, than $\Gamma_r$:

$$\frac{\left|\vec{g}_r\right|^2}{\Gamma_r}=\frac{V_{rr}}{\Gamma_r}=1+2\frac{Q_r}{S}.$$

There is also a sum rule for vectors $\vec{g}_r$:

$$\sum_{r=1}^{N}\left(\vec{g}_r,\vec{g}_r\right)=\sum_{r=1}^{N}\Gamma_r.\qquad\text{(A17)}$$

The branching ratios for the decay of the $r$-th resonance in the $i$-th channel is

$$B_{ri}=\frac{\left|g_{ri}\right|^2}{\left|\vec{g}_r\right|^2}=\frac{\left|g_{ri}\right|^2}{\sum_{k=1}^{M}\left|g_{rk}\right|^2}.\qquad\text{(A18)}$$

For $F = (S - I)/2i$ we have:

$$F_{ij} = \sum_{r=1}^{N} e^{i(\varphi_{ri}+\varphi_{rj})} \frac{|\vec{g}_r|^2 \sqrt{B_{ri}B_{rj}}/2}{E_r - E - i\Gamma_r/2}, \qquad \text{(A19)}$$

where $\varphi_{ri}$ and $\varphi_{rj}$ are real phases of vectors $g_{ri}$ and $g_{rj}$, $g_{rk} = e^{i\varphi_{rk}}|g_{rk}|$.

In a potential fitting procedure there is $2N(M+1)$ real parameters: $E_r, \Gamma_r$, $\vec{g}_r = \vec{g}_r^{\,x} + i\vec{g}_r^{\,y}$. Besides them, the relations (A15) contain $N(N-1)$ elements of real anti-symmetric matrix $U$ $\left(|u_{rl}| \le 1\right)$. (These $u_{ij}$ are 'technical' and do not appear in the final expression for the *S*-matrix.) Not all of these parameters are independent. The $NM$ relations (A6) connect real and imaginary parts of vectors $\vec{g}_r$. Also, the real parts, $\vec{g}_r^{\,x}$, are connected by $N(N+1)/2$ relations (A15). Thus, in total there is $N(M+1)$ free real parameters:

$$\left[2N(M+1) + \frac{N(N-1)}{2}\right] - NM - \frac{N(N+1)}{2} = N(M+1)$$

- the same as in the K-matrix parametrization.

Even though the formulas (A15) look rather sophisticated, the resonance parameters can be determined using a straightforward regular algorithm during, for example, a data fitting procedure; the unitarity of *S* can be checked at each value of *E*. In this Appendix we use the energy variable *E* to avoid unnecessary complications in the formulas, however the algorithm can be rewritten in terms of the variable $s = E^2$, as in section 1 and in one of the Examples in section 3. The codes used for the cases up to 4 resonances (including background) can be provided upon request for data analysis.

For *two resonances* the algorithm is presented in the main text.

For *three resonances,* $N = 3$, matrix *S*(*E*) is

$$S_{ij} = \delta_{ij} + 2i\sqrt{\rho_i\rho_j}\left[\frac{\Gamma_1^0 g_{1i}g_{1j}/2}{\varepsilon_1 - E} + \frac{\Gamma_2^0 g_{2i}g_{2j}/2}{E - \varepsilon_2} + \frac{\Gamma_3^0 g_{3i}g_{3j}/2}{\varepsilon_3 - E}\right], \qquad (i, j = 1,...,M).$$

Matrix *U* relating the real and imaginary parts of vectors $\vec{g}_1$, $\vec{g}_2$, $\vec{g}_3$, is

$$U = \begin{pmatrix} 0 & -\alpha & -\beta \\ \alpha & 0 & -\gamma \\ \beta & \gamma & 0 \end{pmatrix}, \qquad \text{(A20)}$$

i.e.,

$$\vec{g}_1^{\,y} = -\alpha\vec{g}_2^{\,x} - \beta\vec{g}_3^{\,x}, \quad \vec{g}_2^{\,y} = \alpha\vec{g}_1^{\,x} - \gamma\vec{g}_3^{\,x}, \quad \vec{g}_3^{\,y} = \beta\vec{g}_1^{\,x} + \gamma\vec{g}_2^{\,x}.$$

Real parameters $\alpha,\beta,\gamma$ are restricted by the relation $\alpha^2+\beta^2+\gamma^2<1$.
Coefficients in formulas (A16) are:

$$\begin{aligned}
&S=1-\alpha^2-\beta^2-\gamma^2,\\
&Q_1=\alpha^2+\beta^2,\quad Q_2=\alpha^2+\gamma^2,\quad Q_3=\beta^2+\gamma^2,\\
&F_{12}=-\alpha,\quad F_{13}=-\beta,\quad F_{23}=-\gamma;\\
&G_{12}=\beta\gamma,\quad G_{13}=-\alpha\gamma,\quad G_{23}=\alpha\beta.
\end{aligned}\tag{A21}$$

Substituting them in formulas (A16) we obtain the relations needed to construct vectors $\vec{g}_1$, $\vec{g}_2$ and $\vec{g}_3$:

$$\Gamma_1^0\sum_{k=1}^{M}\rho_k(g_{1k}^x)^2=\frac{2}{S^2}[2\alpha\beta\gamma(\varepsilon_3^x-\varepsilon_2^x)-\varepsilon_1^y(1-\gamma^2)^2+\varepsilon_2^y(\alpha^2-\beta^2\gamma^2)+\varepsilon_3^y(\beta^2-\alpha^2\gamma^2)],$$

$$\Gamma_2^0\sum_{k=1}^{M}\rho_k(g_{2k}^x)^2=\frac{2}{S^2}[2\alpha\beta\gamma(\varepsilon_1^x-\varepsilon_3^x)+\varepsilon_1^y(\alpha^2-\beta^2\gamma^2)-\varepsilon_2^y(1-\beta^2)^2+\varepsilon_3^y(\gamma^2-\alpha^2\beta^2)],$$

$$\Gamma_3^0\sum_{k=1}^{M}\rho_k(g_{3k}^x)^2=\frac{2}{S^2}[2\alpha\beta\gamma(\varepsilon_2^x-\varepsilon_1^x)+\varepsilon_1^y(\beta^2-\alpha^2\gamma^2)+\varepsilon_2^y(\gamma^2-\alpha^2\beta^2)-\varepsilon_3^y(1-\alpha^2)^2],$$

$$\begin{aligned}\sqrt{\Gamma_1^0\Gamma_2^0}\sum_{k=1}^{M}\rho_k g_{1k}^x g_{2k}^x=\frac{2}{S^2}\Big\{&\alpha\left[\varepsilon_1^x\left(1-\gamma^2\right)-\varepsilon_2^x\left(1-\beta^2\right)-\varepsilon_3^x\left(\beta^2-\gamma^2\right)\right]+\\
&+\beta\gamma\left[-\varepsilon_1^y\left(1-\gamma^2\right)-\varepsilon_2^y\left(1-\beta^2\right)+\varepsilon_3^y\left(1+\alpha^2\right)\right]\Big\},\end{aligned}$$

$$\begin{aligned}\sqrt{\Gamma_1^0\Gamma_3^0}\sum_{k=1}^{M}\rho_k g_{1k}^x g_{3k}^x=\frac{2}{S^2}\Big\{&\beta\left[\varepsilon_1^x\left(1-\gamma^2\right)-\varepsilon_2^x\left(\alpha^2-\gamma^2\right)-\varepsilon_3^x\left(1-\alpha^2\right)\right]+\\
&+\alpha\gamma\left[\varepsilon_1^y\left(1-\gamma^2\right)-\varepsilon_2^y\left(1+\beta^2\right)+\varepsilon_3^y\left(1-\alpha^2\right)\right]\Big\},\end{aligned}$$

$$\begin{aligned}\sqrt{\Gamma_2^0\Gamma_3^0}\sum_{k=1}^{M}\rho_k g_{2k}^x g_{3k}^x=\frac{2}{S^2}\Big\{&\gamma\left[-\varepsilon_1^x\left(\alpha^2-\beta^2\right)+\varepsilon_2^x\left(1-\beta^2\right)-\varepsilon_3^x\left(1-\alpha^2\right)\right]+\\
&+\alpha\beta\left[\varepsilon_1^y\left(1+\gamma^2\right)-\varepsilon_2^y\left(1-\beta^2\right)-\varepsilon_3^y\left(1-\alpha^2\right)\right]\Big\}.\end{aligned}$$

The widths are

$$\Gamma_1 = \Gamma_1(\hat{\varepsilon}_1) = \frac{1-\alpha^2-\beta^2-\gamma^2}{2\left(1+\alpha^2+\beta^2-\gamma^2\right)}\Gamma_1^0 \sum_{k=1}^{M} \rho_k(\hat{\varepsilon}_1)\left|g_{1k}\right|^2,$$

$$\Gamma_2 = \Gamma_2(\hat{\varepsilon}_2) = \frac{1-\alpha^2-\beta^2-\gamma^2}{2\left(1+\alpha^2-\beta^2+\gamma^2\right)}\Gamma_2^0 \sum_{k=1}^{M} \rho_k(\hat{\varepsilon}_2)\left|g_{2k}\right|^2, \qquad \text{(A22)}$$

$$\Gamma_3 = \Gamma_3(\hat{\varepsilon}_3) = \frac{1-\alpha^2-\beta^2-\gamma^2}{2\left(1-\alpha^2+\beta^2+\gamma^2\right)}\Gamma_3^0 \sum_{k=1}^{M} \rho_k(\hat{\varepsilon}_3)\left|g_{3k}\right|^2,$$

where $\hat{\varepsilon}_i$ are zeroes of real parts of denominators in the corresponding BW terms.

Very briefly describe the case of *four resonances*, $N = 4$. Matrix $U$ is

$$U = \begin{pmatrix} 0 & -\alpha & -\beta & -\delta \\ \alpha & 0 & -\gamma & -\eta \\ \beta & \gamma & 0 & -\omega \\ \delta & \eta & \omega & 0 \end{pmatrix} \qquad \text{(A23)}$$

thus,

$$\begin{aligned} \vec{g}_1^{\,y} &= -\alpha\vec{g}_2^{\,x} - \beta\vec{g}_3^{\,x} - \delta\vec{g}_4^{\,x}, \\ \vec{g}_2^{\,y} &= \alpha\vec{g}_1^{\,x} - \gamma\vec{g}_3^{\,x} - \eta\vec{g}_4^{\,x}, \\ \vec{g}_3^{\,y} &= \beta\vec{g}_1^{\,x} + \gamma\vec{g}_2^{\,x} - \omega\vec{g}_4^{\,x}, \\ \vec{g}_4^{\,y} &= \delta\vec{g}_1^{\,x} + \eta\vec{g}_2^{\,x} + \omega\vec{g}_3^{\,x}. \end{aligned} \qquad \text{(A24)}$$

The lengths of vectors $\vec{g}_1^{\,x}$, $\vec{g}_2^{\,x}$, $\vec{g}_3^{\,x}$, $\vec{g}_4^{\,x}$ and the angles between them are determined by the scalar products (A15) for $N = 4$. From that we have for the coefficients in formulas (A16):

$S = 1-\alpha^2-\beta^2-\delta^2-\gamma^2-\eta^2-\omega^2+\tilde{D}^2,$

$Q_1 = \alpha^2+\beta^2+\delta^2-\tilde{D}^2, \quad Q_2 = \alpha^2+\gamma^2+\eta^2-\tilde{D}^2,$

$Q_3 = \beta^2+\gamma^2+\omega^2-\tilde{D}^2, \quad Q_4 = \delta^2+\eta^2+\omega^2-\tilde{D}^2,$

$F_{21} = \alpha-\omega\tilde{D}, \quad F_{31} = \beta+\eta\tilde{D}, \quad F_{41} = \delta-\gamma\tilde{D}$

$F_{32} = \gamma-\delta\tilde{D}, \quad F_{42} = \eta+\beta\tilde{D}, \quad F_{43} = \omega-\alpha\tilde{D},$

$F_{12} = -\alpha+\omega\tilde{D}, \quad F_{13} = -\beta-\eta\tilde{D}, \quad F_{14} = -\delta+\gamma\tilde{D},$

$F_{23} = -\gamma+\delta\tilde{D}, \quad F_{42} = -\eta-\beta\tilde{D}, \quad F_{34} = -\omega+\alpha\tilde{D}.$

$G_{21} = \beta\gamma+\delta\eta, \quad G_{12} = \beta\gamma+\delta\eta, \quad G_{31} = -\alpha\gamma+\delta\omega,$

$G_{41} = -\alpha\eta-\beta\omega, \quad G_{32} = \alpha\beta+\eta\omega,$

$G_{42} = \alpha\delta - \gamma\omega$, $G_{43} = \beta\delta + \gamma\eta$,
where $\tilde{D} = \alpha\omega - \beta\eta + \gamma\delta$.

Formulas (A15) give the scalar products - as an example we show two of them:

$$\Gamma_1^0 \sum_{k=1}^{M} \rho_k (g_{1k}^x)^2 = \frac{2}{S^2} \cdot \Big\{ -2\left(\beta\gamma + \delta\eta\right)\left(\alpha - \omega\tilde{D}\right) \cdot \varepsilon_2^x + 2\left(\alpha\gamma - \delta\omega\right)\left(\beta + \eta\tilde{D}\right) \cdot \varepsilon_3^x +$$
$$+ 2\left(\alpha\eta + \beta\omega\right)\left(\delta - \gamma\tilde{D}\right) \cdot \varepsilon_4^x - \left(1 - \gamma^2 - \eta^2 - \omega^2\right)^2 \cdot \varepsilon_1^y +$$
$$+ \left[\left(\alpha - \omega\tilde{D}\right)^2 - \left(\beta\gamma + \delta\eta\right)^2\right] \cdot \varepsilon_2^y + \left[\left(\beta + \eta\tilde{D}\right)^2 - \left(\alpha\gamma - \delta\omega\right)^2\right] \cdot \varepsilon_3^y +$$
$$+ \left[\left(\delta - \gamma\tilde{D}\right)^2 - \left(\alpha\eta + \beta\omega\right)^2\right] \cdot \varepsilon_4^y \Big\},$$

$$\sqrt{\Gamma_2^0 \Gamma_1^0} \sum_{k=1}^{M} \rho_k g_{2k}^x g_{1k}^x = \frac{2}{S^2} \Big\{ \left(1 - \gamma^2 - \eta^2 - \omega^2\right)(\alpha - \omega\tilde{D})\, \varepsilon_1^x -$$
$$- \left(1 - \beta^2 - \delta^2 - \omega^2\right)(\alpha - \omega\tilde{D})\, \varepsilon_2^x -$$
$$- \left[(-\alpha\gamma + \delta\omega)(\gamma - \delta\tilde{D}) + (\alpha\beta + \eta\omega)(\beta + \eta\tilde{D})\right] \varepsilon_3^x -$$
$$- \left[(-\alpha\eta - \beta\omega)(\eta + \beta\tilde{D}) + (\alpha\delta - \gamma\omega)(\delta - \gamma\tilde{D})\right] \varepsilon_4^x -$$
$$- (\beta\gamma + \delta\eta)\left(1 - \gamma^2 - \eta^2 - \omega^2\right) \varepsilon_1^y -$$
$$- (\beta\gamma + \delta\eta)\left(1 - \beta^2 - \delta^2 - \omega^2\right) \varepsilon_2^y +$$
$$+ \left[(\beta + \eta\tilde{D})(\gamma - \delta\tilde{D}) - (-\alpha\gamma + \delta\omega)(\alpha\beta + \eta\omega)\right] \varepsilon_3^y +$$
$$+ \left[(\delta - \gamma\tilde{D})(\eta + \beta\tilde{D}) - (-\alpha\eta - \beta\omega)(\alpha\delta - \gamma\omega)\right] \varepsilon_4^y \Big\}.$$